\documentstyle[epsfig,floats,amssymb,aps,prl,twocolumn]{revtex}

\begin{document}

\draft

\title{Dressed States of a two component Bose-Einstein Condensate.}

\author{P.~B.~Blakie and R.~J.~Ballagh}
\address{Department of Physics, University of Otago, P.~O.~Box~56, Dunedin,
New
Zealand.}

\author{C. W. Gardiner}
\address{Department of Physics, Victoria University of Wellington,
Wellington, New Zealand.}

\date{\today}
\wideabs{
\maketitle

\begin{abstract}
A condensate with two internal states coupled by external electromagnetic
radiation, is described by coupled Gross Pitaevskii equations, whose
eigenstates are analogous to the dressed states of quantum optics. We solve
for these eigenstates numerically in the case of one spatial dimension, and
explore their properties as a function of  system parameters. In contrast to
the quantum optical case, the condensate dressed states exhibit spatial
behaviour which depends on the system parameters, and can be manipulated by
changing the cw external field.

\medskip
\noindent PACS number(s) 03.75.Fi, 05.30.Jp 
\end{abstract}
}
\section{Introduction}
Recent experimental work with multiple species Bose Einstein 
condensates (BEC) \cite{binexpt1,binexpt4} has motivated theoretical
analysis 
of their wavefunctions \cite{binwf1,binwf2} and excitations 
\cite{binexcit1,binexcit3,binexcit2}.  The primary tools used to create and
investigate 
these multiple condensates are external microwave and radiofrequency 
radiation fields, and the combined system is described by a set of 
coupled Gross Pitaevskii equations (GPE).  In quantum optics the 
eigenstates of the full system of the (single) atom plus field are 
called {\em dressed states} (see \cite{atomphoton}) and have proved
invaluable as 
calculational and interpretational tools.  In this paper we extend 
this concept to the eigenstates of the coupled GPE, which we 
shall call {\em condensate dressed states}.
 
We consider in detail a condensate with two internal states 
$|1\rangle$ and $|2\rangle$.  We solve for the eigenstates of this 
system numerically in the case of one spatial dimension, and with 
plane wave uniform intensity radiation fields.  The major new feature 
that occurs in condensate dressed states is the spatial dependence of 
the wave functions, which can prove significant even when the external 
field is a uniform plane wave.

We explore the properties of the condensate dressed states as a 
function of system parameters, and present results representing broad 
classes of the possible behaviour.  We begin by considering the most 
general properties of the condensate dressed states including their 
symmetries.  We show that in the simplest case of identical traps for 
each component and identical collisional interactions, both components 
have identical spatial behaviour.  However, with non identical traps, 
the two components may have markedly different spatial character, and 
we examine the dependence of these shapes on the trap parameters, the 
collisional parameters, and the external field.  We show that the 
condensate's spatial shapes can be manipulated by changing the 
external field strength or detuning.

\section{Formulation}

The single particle state, $|\Psi\rangle$, of a condensate in a 
superposition of the two internal states $|1\rangle$ and $|2\rangle$ 
can be written
\begin{equation}
 \label{dresssup}
 |\Psi\rangle = \phi_1({\mathbf r},t)\,|1\rangle+\phi_2({\mathbf 
 r},t)\,|2\rangle,
\end{equation}
where $\phi_i({\mathbf r},t)$ is the centre of mass meanfield 
wavefunction for a particle in state $|i\rangle$.  Under the influence 
of an electromagnetic coupling the wavefunctions associated with each 
component evolve according to the coupled Gross Pitaevskii equations 
(GPE)

\begin{eqnarray}
\label{GPE}
i\,{\partial \phi_1\over \partial t} & = & -\nabla^2\phi_1+V_1({\mathbf 
r})\phi_1+\left [w_{11}|\phi_1|^2+w_{12}|\phi_2|^2\right ]\,\phi_1\\
&&-{\Omega 
\over 2}\phi_2, 
\cr
\smallskip \cr
i\,{\partial \phi_2 \over \partial t} & = & -\nabla^2\phi_2+V_2({\mathbf 
r})\phi_1+\left [w_{22}|\phi_2|^2+w_{12}|\phi_1|^2\right ]\,\phi_2\\
&&-{\Omega
\over 
2}\phi_1+\delta_L\phi_2, 
\end{eqnarray}
which describe either a 1-photon \cite{outcup} or 2-photon Raman 
\cite{vort1,vort2,coupling} transition.  In Eq. (\ref{GPE}) the scaling is 
chosen as in \cite{outcup}, and $\Omega$ and $\delta_L$ are the Rabi 
frequency and the detuning for the transition.  The trapping 
potentials $V_1({\mathbf r})$ and $V_2({\mathbf r})$ for atoms in 
internal states $|1\rangle$ and $|2\rangle$ respectively, may describe 
any static potential, but here we restrict our attention to the case 
where they are each harmonic, but possibly with different spring 
constants and trap centres.  The eigenvalue, $\mu$, at $T=0$ can be 
identified as the chemical potential of the system.  The quantities 
$w_{11}$, $w_{22}$ and $w_{12}$ represent the strength of the two 
intra- and the inter-species interactions, and are proportional to the 
total number of atoms within the condensed system and the respective 
scattering lengths.  The state $|\Psi\rangle$ is normalised to unity, 
so that the fractional population $n_i$, of the state $|i\rangle$, is given 
by the $n_i = \int|\psi_i|^2d^3r$.

We look for stationary solutions to Eq. (\ref{GPE}) of the form 
$\phi_{1}({\mathbf r},t)=\psi_1({\mathbf r})e^{-i\mu t}$, 
$\phi_{2}({\mathbf r},t)=\psi_2({\mathbf r})e^{-i\mu t}$ 
which gives rise to the time independent coupled Gross Pitaevskii 
Equations

\begin{eqnarray}
\label{tiGPE}
\mu\,\psi_1 & = & -\nabla^2\psi_1+V_1({\mathbf 
r})\psi_1+\left [w_{11}|\psi_1|^2+w_{12}|\psi_2|^2\right ]\,\psi_1\\
&&-{\Omega
\over 2}\psi_2, 
\cr
\smallskip \cr
\mu\,\psi_2 & = & -\nabla^2\psi_2+V_2({\mathbf 
r})\psi_1+\left [w_{22}|\psi_2|^2+w_{12}|\psi_1|^2\right]\,\psi_2-\\
&&{\Omega
\over 
2}\psi_1+\delta_L\psi_2, 
\end{eqnarray}

In general this equation must be solved numerically, and we have used 
an optimisation technique (which will be presented elsewhere 
\cite{optimisation}) to find solutions.  We present and discuss 

solutions which represent broad classes of possible behaviour, which 
we discuss in the next two sections.

\section{General Properties}

Many properties of the dressed condensate turn out to be analogous to 
those of the familiar dressed states of quantum optics 
\cite{atomphoton}, however a key difference is the stationary 
solutions of Eq. (\ref{tiGPE}) are spatially dependent.  For simplicity 
we shall concentrate in this paper on solutions $\psi_1$ and $\psi_2$ 
of Eq. (\ref{tiGPE}), that have no nodes.  In this case two eigenstates 
can be found, which we label $|\Psi_{\pm}\rangle$ 
and denote the corresponding eigenvalues as $\mu_{\pm}$, where 
$\mu_+>\mu_-$.  Apart from constant overall phase factors, the 
mathematical form of these states is
\begin{eqnarray}
 \label{dressedpair}
 |\Psi_+\rangle & = & \psi_{1+}({\mathbf r})\,|1\rangle - 
\psi_{2+}({\mathbf 
 r})\,|2\rangle, \cr
 \smallskip \cr
 |\Psi_-\rangle & = & \psi_{1-}({\mathbf r})\,|1\rangle + 
\psi_{2-}({\mathbf 
 r})\,|2\rangle,
\end{eqnarray}
where the component wavefunctions $\psi_{1\pm}$ and $\psi_{2\pm}$ are
positive real 
functions.
\begin{figure}
\begin{center}
\epsfbox{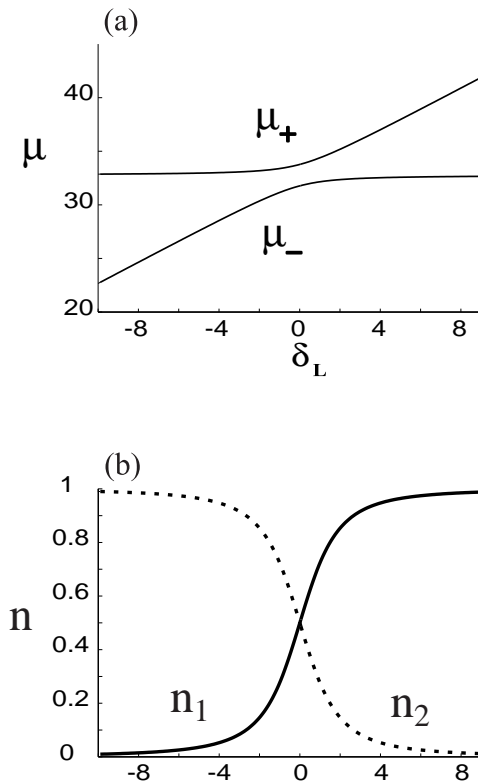}
\end{center}
\caption{Properties of condensate dressed states in frequency 
scans.  (a) Eigenvalues $\mu_+$ and $\mu_-$.  (b) Populations 
$n_1$ (solid line) and $n_2$ (dashed line) of the internal states 
$|1\rangle$ and $|2\rangle$.  Parameters are $V_1 = V_2 = x^2/4$, 
$w_{11}=w_{12}=w_{22}=500$ and $\Omega=2$.}
\label{fig:genprops}
\end{figure}
Scanning the detuning of the electromagnetic field from the far red 
through to the far blue and solving for $|\Psi_{\pm}\rangle$ and 
$\mu_{\pm}$ at each point reveals that the eigenvalues observe an 
avoided crossing (see Fig.  \ref{fig:genprops}(a)), very similar to 
that seen in quantum optics.  Here it is associated with a resonance 
in which the component populations are near equal (typically when 
$|\delta_{L}|<\Omega$), as shown in Fig.  \ref{fig:genprops} (b). On either 
side of this resonance, as $|\delta_L|$ increases, the dressed states 
approach a single component configuration (i.e the dressed state is 
almost entirely in one internal state).

The states of Eq.  (\ref{dressedpair}) exhibit certain symmetry 
properties, which relate the solution $|\Psi\rangle = \pmatrix{\psi_1 \cr 
\psi_2}$ for field parameters $\{\Omega,\delta_L\}$ to another 
solution $|\bar{\Psi}\rangle = \pmatrix{\bar{\psi}_1 \cr 
\bar{\psi}_2}$ for $\{\bar{\Omega},\bar{\delta}_L\}$. For the case 
of identical trapping potentials ($V_1=V_2$) and equal collisional strengths

($w_{11}=w_{12}=w_{22}$), the symmetries are the same as those of the 
quantum optics dressed states, namely if the field is related by the
rotation matrix 
$R(\theta)$ according to\begin{equation}
 \pmatrix{\bar{\Omega} \cr \bar{\delta}_L} = R(\theta)\pmatrix{\Omega \cr
\delta_L},
\end{equation}
where
\begin{equation}
 R(\theta) = \pmatrix{\cos\theta & \sin\theta \cr -\sin\theta & 
 \cos\theta}.
\end{equation}
The corresponding dressed eigenvector is given by
\begin{equation}
 \pmatrix{\bar{\psi_1} \cr \bar{\psi_2}} = R\left ({ \theta \over 
 {2}}\right )\pmatrix{\psi_1\cr 
 \psi_2},
\end{equation}
with eigenvalue 
\begin{equation}
 \bar{\mu} = \mu - \delta_L\sin^2{\theta\over 2} -{\Omega\over 2} \sin
\theta.
\end{equation}

For $w_{11}=w_{22}\ne w_{12}$ the symmetry is reduced to the single 
transformation
\begin{eqnarray}
\label{symmetry}
\bar{\psi_1} & =  \psi_2, \qquad \bar{\psi_2} & =  \psi_1. \cr
\bar{\mu} & =  \mu-\delta_L, \qquad\bar{\delta}_L & =  -\delta_L.
\end{eqnarray}

\section{Spatial Characteristics of Dressed States}
Perhaps the main interest of the dressed states is that they provide a 
simple means for manipulating the spatial shape of the condensate 
components.  However this is very dependent on the relative trap 
potentials and collisional parameters.

In the simplest case, where the trapping potentials are identical 
($V_{1}=V_{2}$) and 
collisional coupling coefficients are all equal, ($w_{ij} = w$), no 
difference in spatial characteristics of the components is possible. 
It is easy to show the that eigenstates $|\Psi_{\pm}\rangle$ can be written 
in terms of the eigenfunction $\psi_o$ of the uncoupled 
one component GPE
\begin{equation}
 \mu_o\,\psi_o({\mathbf r}) = -\nabla^2\psi_o+V_1\psi_o({\mathbf r})
 +w|\psi_o|^2\,\psi_o({\mathbf r}),
\end{equation}
and take the general form
\begin{equation}
|\Psi_\pm\rangle  =  c_{1\pm}\psi_o({\mathbf r})\,|1\rangle \mp 
c_{2\pm}\psi_o({\mathbf r})\,|2\rangle, 
\end{equation}
with eigenvalues
\begin{equation} 
\mu_\pm  =  \mu_o + {1\over 2} \left ( \delta_L \pm 
\sqrt{\delta_L^2+\Omega^2} \right ),
\end{equation}
where
\begin{eqnarray}
 c_{1\pm}^2 & = & {1\over 2} \left 
 (1\mp{\delta_L\over\sqrt{\delta_L^2+\Omega^2}}\right), \cr
 \smallskip\cr
  c_{2\pm} & = & \mp\sqrt{1-c_{1\pm}^2}.
\end{eqnarray}
These amplitudes ($c_{j\pm}$) and eigenvalues have precisely the same 
dependence on field parameters $\Omega$ and $\delta_L$ as for the quantum 
optics dressed state. 

\subsection{ Effect of Different Relative Trap Potentials}
\begin{figure}
\begin{center}
\epsfbox{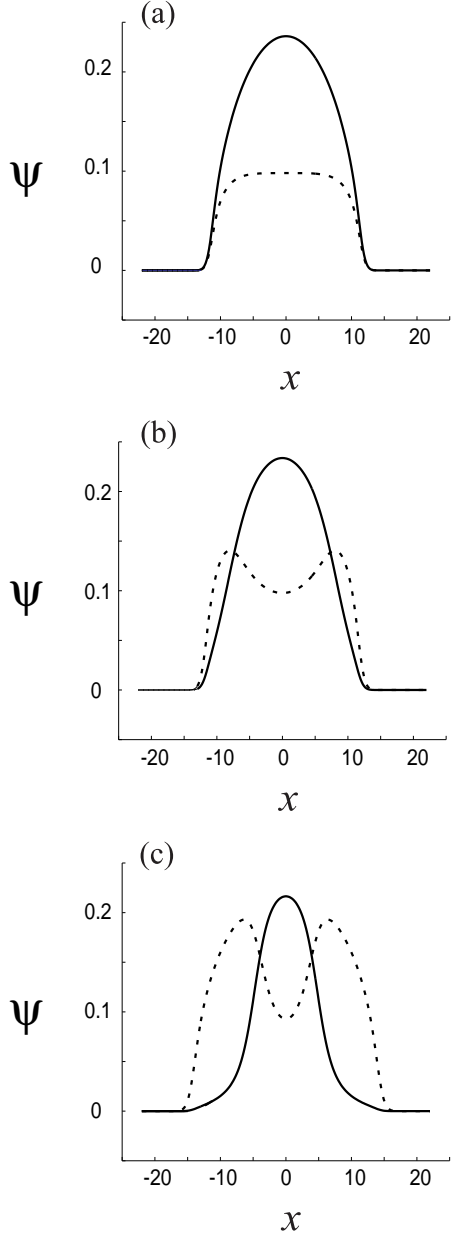}
\caption{Effect of relative spring constant on component 
wavefunctions of dressed states for state $|1\rangle$ (solid line) 
and state $|2\rangle$ (dashed line).  (a) $k=0.95$, (b) $k=0.85$, 
(c) $k=0.5$.  Parameters are as in Fig.  \ref{fig:genprops}, 
except $V_2=k x^2/4$ and $\delta_L=2$.}
\label{fig:spring}
\end{center}
\end{figure}

\begin{figure}
\begin{center}
\epsfbox{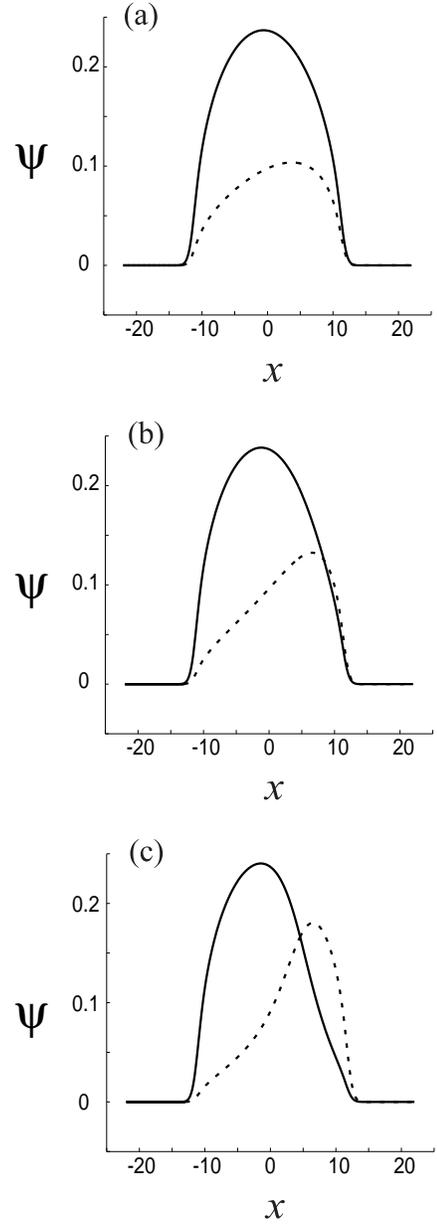}
\caption{Effect of trap offset on component wavefunctions. (a) 
$x_o=0.2$, (b) $x_o=0.5$, (c) $x_o=1$. 
Parameters are as in Fig.  
\ref{fig:genprops}, except $V_2=(x-x_o)^2/4$ and $\delta_L=2$.}
\label{fig:offset}
\end{center}
\end{figure}

For simplicity, we will consider only harmonic traps with $V_1 = 
x^2/4$ and $V_2 = k(x-x_o)^2/4$, which allows condensate $|2\rangle$ 
to have a different relative spring constant ($k$) and an offset 
centre ($x_o$).  This arises in the JILA experiment, for example, 
because of the different magnetic moment of the two components and the 
effect of gravity (see \cite{binexpt3}).  No exact analytic solutions are
possible in this 
case, but the representative behaviour is shown in numerical solutions 
presented in Figs. \ref{fig:spring} and \ref{fig:offset}.  In Fig. 
\ref{fig:spring}(a) we see that even a small difference in the relative
spring 
constant (5\%) can give rise to a significant difference in the 
component wavefunctions, and as $k$ is further increased a clear phase 
separation is observed (Figs. \ref{fig:spring}(b)-(c)), whereby one 
species is excluded from the region where the other species has high 
density.

Offsets between the two trapping potentials also cause spatial 
reshaping of the two components, as can be seen in Fig. 
\ref{fig:offset}, where $x_o$ is successively increased.  In Fig. 
\ref{fig:offset}(a), where $x_o$ is approximately 1\% of the 
condensate size, significant deformation of component 2 is seen, and 
when $x_o$ is increased to 1 (i.e 10\% of condensate size) phase separation
occurs (Figs. 
\ref{fig:offset}(b)-(c)).

\subsection{Rabi Frequency Dependence}
\begin{figure}
\begin{center}
\epsfbox{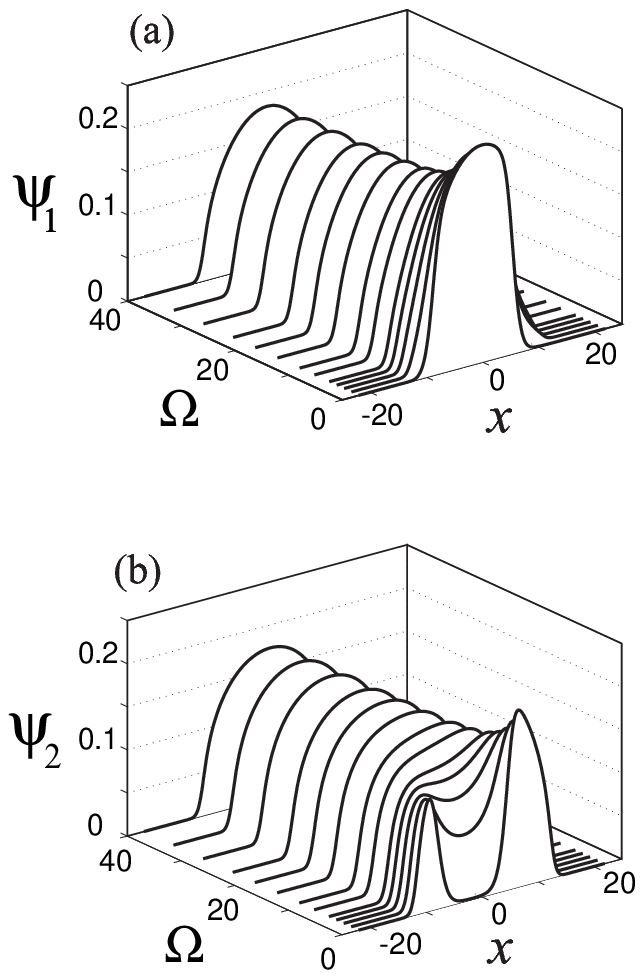}
\caption{Effect of changing Rabi frequency on component 
wavefunctions.  (a) component $|1\rangle$.  (b) component 
$|2\rangle$.  The trapping potentials are $V_1=x^2/4$, 
$V_2=k(x-x_o)^2/4$ where $k=0.8$ and $x_o=0.5$, the detuning is 
 $\delta_L=2$ and the $w_{ij}=500$.}
\label{fig:omega}
\end{center}
\end{figure}
A scan of the component wavefunction shapes as $\Omega$ is varied is 
shown in Fig. \ref{fig:omega}, which illustrates the control over the 
condensate profiles that is afforded by simply altering the strength 
of the electromagnetic coupling field.  Two limiting regimes can be 
seen.  At low fields ($\Omega\ll|\delta_L|$) the components have 
distinct spatial shapes, determined by different trap potentials and 

collisional interactions.  On the other hand, in the large field 
regime ($\Omega\gg|\delta_L$) $\Omega$ becomes the most significant 
coupling between components and suppresses the spatial differences 
between the component wavefunctions.  In particular, we notice that at 
large Rabi frequency phase separation between the condensates will 
be suppressed.

\subsection{Effect of Different Relative Collisional Interactions}

All of the previous results have been given for the case of equal 
collisional interactions ($w_{11}=w_{12}=w_{22}$).  Although this is a 
good approximation to the case of Rubidium, larger variations can be 
expected for other atomic species, and it may even prove possible to 
manipulate the relative scattering lengths (e.g see \cite{inter1}).  
It has previously been shown that the extent of phase separation of 
binary condensates in the absence of electromagnetic coupling depends 
on the relative collisional interactions \cite{binwf1}.  Here we find 
analogous features arising in our dressed state solutions.

\begin{figure}
\begin{center}
\epsfbox{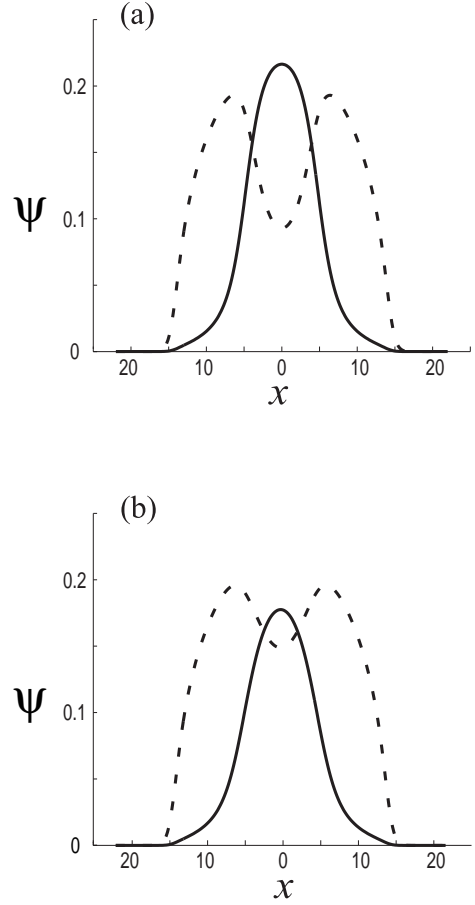}
\caption{Effect of changing $w_{11}$ relative to $w_{12}$ and 
$w_{22}$ on the component wavefunctions, with state $|1\rangle$ 
(solid line) and state $|2\rangle$ (dashed line).  (a) $w_{11}$ = 
500.  (b) $w_{11}$ = 550.  Other parameters are as in Fig.  
\ref{fig:spring} (c).}
\label{fig:intrachange}
\end{center}
\end{figure}

In Fig. \ref{fig:intrachange}, we illustrate the effect of increasing 
$w_{11}$ relative to $w_{22}$.  This has the effect of increasing the 
self energy of component 1, thus making it less favoured, and 
resulting in the density of component one being reduced in comparison 
to component 2.  Similarly, decreasing $w_{11}$ makes component 1 
energetically favoured, thus causing an increase in its density.

\begin{figure}
\begin{center}
\epsfbox{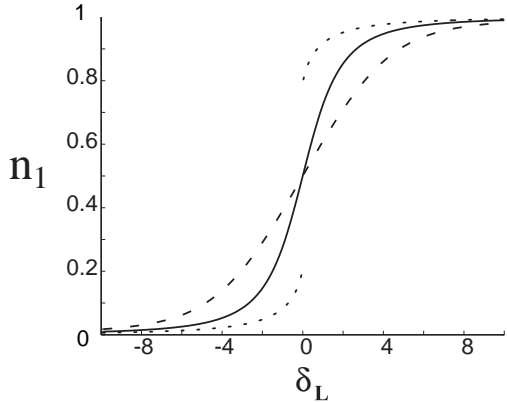}
\caption{Component population ($n_1$) in frequency scans with different 
interparticle collisional interactions. Solid line $w_{12}=500$, 
dashed line $w_{12}=450$ and dotted line $w_{12}=550$. Other 
parameters are $V_1=V_2=x^2/4$, $\delta_L=2$, $\Omega=2$, and 
$w_{ii}=500$.}
\label{fig:adiabatic}
\end{center}
\end{figure}
  
The cross coupling term $w_{12}$ mediates the interaction between the 
two components.  When $w_{12}$ is larger than $w_{11}$ and $w_{22}$, a 
competitive interaction occurs, so that the larger component in any 
region is favoured at the expense of the other component, thereby 
resulting in density differences being enhanced.  Correspondingly, 
when $w_{12}$ is smaller than $w_{11}$ and $w_{22}$, it becomes 
energetically favourable for the components to coexist, and density 
differences are reduced.  The influence that $w_{12}$ exerts on mixing 
can be clearly shown by examining how the total populations of the two 
internal states change as the field frequency is scanned.  In Fig.  
\ref{fig:adiabatic} we see that for the case where all collisional 
rates are equal (solid line) the total population ($n_1$) in state 
$|1\rangle$ changes from 0.3 to 0.7 over the range 
$-1.0<\delta_L<1.0$.  However, to see the same change in $n_1$ when 
$w_{12}$ is decreased by $10\%$ (dotted line), the detuning range has 
to be extended to $-2.0<\delta_L<2.0$.  Reducing $w_{12}$ is thus seen 
to favour mixing of the components.  On the other hand, when $w_{12}$ 
is larger than $w_{ii}$, mixing is unfavourable and one state will 
usually dominate the other.  This effect can be seen with the dashed 
line in Fig.  \ref{fig:adiabatic}, where when the frequency scan 
passes through zero, a very abrupt reversal in the dominant state 
occurs.
\subsection{Conclusion}
We have given a preliminary investigation of condensate dressed 
states, and explored the similarities and differences to the familiar 
dressed states of quantum optics.  The major new feature that occurs 
is that the two components of a condensate dressed state may exhibit 
distinctly different spatial shapes, and we have examined how these 
shapes depend on the various system parameters.  We have considered 
only the simplest case where the component wavefunctions have no 
nodes, and the external field is a uniform intensity plane wave, and 
shown that even in this case the shape of the component wavefunctions 
can be manipulated by varying the field parameters.

Recent investigations \cite{vort1,vort2} have shown that when 
condensates components are coupled by a spatially varying field, an 
adiabatic change of detuning may be used to generate an excited 
condensate state from the ground state.  We can generalise our dressed 
state analysis to include the possibility of a radiation field with 
spatial structure, in which case the individual component wave 
functions may have different numbers of nodes.  In this context, the 
dressed states may be useful to help predict the outcome of adiabatic 
passage, including the possibility of crossing to a branch of 
different symmetry.

This work was supported 
by Marsden Grant PVT 603.

\end{document}